\documentclass{elsart}
\usepackage{epsfig}
\usepackage{upgreek}
\journal{Nuclear Instruments and Methods A}
\begin{document}
\begin{frontmatter}
\title{Timing performance of small cell 3D silicon detectors \thanksref{RD50}}
\thanks[RD50]{Work performed in the framework of the CERN-RD50 collaboration.}
\author{G.\,Kramberger$^{a}$, V.\,Cindro$^{a}$, D. Flores$^{b}$,}
\author{S.\,Hidalgo$^{b}$, B.\,Hiti$^{a}$, M. Manna$^{b}$, I.\,Mandi\' c$^{a}$, M.\,Miku\v z$^{a,c}$,}
\author{ D.\,Quirion$^{b}$, G. Pellegrini$^{b}$, M. Zavrtanik$^{a}$}
\address {
$^a$ Jo\v zef Stefan Institute, Jamova 39, SI-1000 Ljubljana, Slovenia\\
$^b$ Centro Nacional de Microelectr\'{o}nica (IMB-CNM-CSIC), Barcelona 08193, Spain \\
$^c$ University of Ljubljana, Faculty of Mathematics and Physics, Jadranska 19, SI-1000 Ljubljana, Slovenia \\}
\thanks[hvala]{Corresponding author; Address:
Jo\v zef Stefan Institute, Jamova 39, SI-1000 Ljubljana,
Slovenia. Tel: (+386) 1 477 3512, fax: (+386) 1 477 3166,
e-mail: Gregor.Kramberger@ijs.si}     
\begin{abstract}
A silicon 3D detector with a single cell of $50 \times 50$ $\upmu$m$^2$ was produced and evaluated for
timing applications. The measurements of time resolution were performed for $^{90}$Sr electrons with 
dedicated electronics used also for determining time resolution of Low Gain Avalanche Detectors (LGADs). 
The measurements were compared to those with LGADs and also simulations. The studies showed that the  
dominant contribution to the timing resolution comes from the time walk originating from different 
induced current shapes for hits over the cell area. This contribution
decreases with higher bias voltages, lower temperatures and smaller cell sizes. It is around
30 ps for a 3D detector of $50 \times 50$ $\upmu$m$^2$ cell at 150 V and  -20$^\circ$C, which is comparable 
to the time walk due to Landau fluctuations in LGADs. It even improves for inclined 
tracks and larger pads composed of multiple cells. A good agreement between 
measurements and simulations was obtained, thus validating the simulation results.   
\end{abstract}
\end{frontmatter}
\section{Introduction}

The choice of solid state timing detectors to be used at large experiments CMS \cite{CMS-TD}
and ATLAS \cite{HGTD} after the luminosity upgrade of the LHC (HL-LHC) are presently thin Low Gain Avalanche Detectors 
(LGAD) \cite{GP2014}. 
They rely on charge multiplication in the so called gain layer, a heavily doped 1-2 $\upmu$m thick $p^+$ 
layer sandwiched between the $n^{++}$ implant and the $p$ bulk. Gains of $>10$ allow efficient 
operation of thin detectors ($\sim 50$ $\upmu$m) required for superior time
resolution \cite{LGAD-Thin} of around 30 ps per detector layer \cite{LGAD-Timing}. 

However, the gain degrades 
with irradiation \cite{GK2015,GK2018}. The high gain of LGADs can be maintained at equivalent fluences 
below $10^{15}$ cm$^{-2}$, where their performance has been demonstrated to fulfill the HL-LHC 
requirements \cite{HGTD-TP}. At fluences above that, particularly of charged hadrons, the operation 
requires extremely high bias voltages of $>\approx 600$ V to achieve small gain factors (of only few)
\cite{GK2018}. With the loss of gain and consequent decrease of signal-to-noise ratio ($S/N$) the 
time resolution and detection efficiency of thin LGADs degrades. Operation of LGADs 
close to breakdown voltage poses a risk and so far there is also no running experience over years of operation. 

Another problem of LGADs are special junction termination structures used to isolate pixels/pads and prevent early breakdowns, which lead 
to a region between pixels/pads without the gain \cite{LGADpixelISo}.
This region has typically a width of the order of 40-100 $\upmu$m, which even for relatively large pads of $1.3\times1.3$ mm$^2$ 
leads to a significant reduction of a fill factor of up to 13\%. More importantly, this prevents using LGADs with 
smaller pads, which would be required for a smaller pad/pixel capacitance.
The fill factor can be resolved by using so called inverse LGADs (iLGAD) \cite{iLGAD}, which however require even more  
complex processing. A problem of decreasing gain with irradiation is only moderately improved by the  
carbon co-implantation in gain layer \cite{NC2018-1} or replacement of boron with gallium \cite{NC2018-1,GK2018-1}. As a result of 
the above mentioned limitations of LGADs alternatives are sought. 

Recent results with 3D detectors produced 
by CNM\footnote{Centro Nacional de Microelectr\'{o}nica (IMB-CNM-CSIC), Barcelona, Spain}, 
which have a cell size compatible with the RD53 readout chip\cite{RD53}, both in test
beam \cite{JLange2018} and with $^{90}$Sr electrons \cite{RD50-CNM3D-1,RD50-CNM3D-2}, showed only small degradation
of charge collection with fluence over the entire HL-LHC fluence range, with most probable signal $>16000$ e for 230 $\upmu$m thick detector. Efficient charge collection together with small drift distances, which lead to short induced current pulses, offer a possibility for their 
use also in timing applications.

The aim of this paper is an investigation of small cell 3D detectors 
in timing applications. Simulations and measurements with $^{90}$Sr 
electrons will be discussed in the work.

\section{Time resolution} 

The time resolution ($\sigma_t$) of a detector is to a large extent given by 
\begin{eqnarray}
\sigma_t^2 & = & \sigma_{j}^2 + \sigma_{tw}^2 \\ 
\sigma_{j} & = &N/(dV/dt) \sim t_p/(S/N) \quad ,
\label{eq:TimeRes}
\end{eqnarray}  
where $\sigma_{j}$ is the jitter contribution determined by the rise of the signal at the output of the 
amplifier $dV/dt$ and noise level $N$ ($t_p$ is peaking time of electronics and $S/N$ signal to noise ratio) 
and $\sigma_{tw}$ is the time walk contribution. The latter is usually minimized by using 
Constant Fraction Discrimination or determining time of the signal crossing fixed threshold and its duration 
over that threshold. These two techniques eliminate the difference in signal height arising from the amount of 
deposited charge in the sensor, but not the differences in signal shapes. The shape of the signal is mainly affected 
by the differences in drift paths (depending on the hit position hit positions inside the pixel cell) 
of generated carriers, which drift with different drift velocities in different weighting fields. Fluctuations 
in ionization rates along the track path (Landau fluctuations) add to the differences in pulse shapes. 
These two contributions usually dominate the time walk.

\begin{figure}[!hbt]
\begin{center}
\epsfig{file=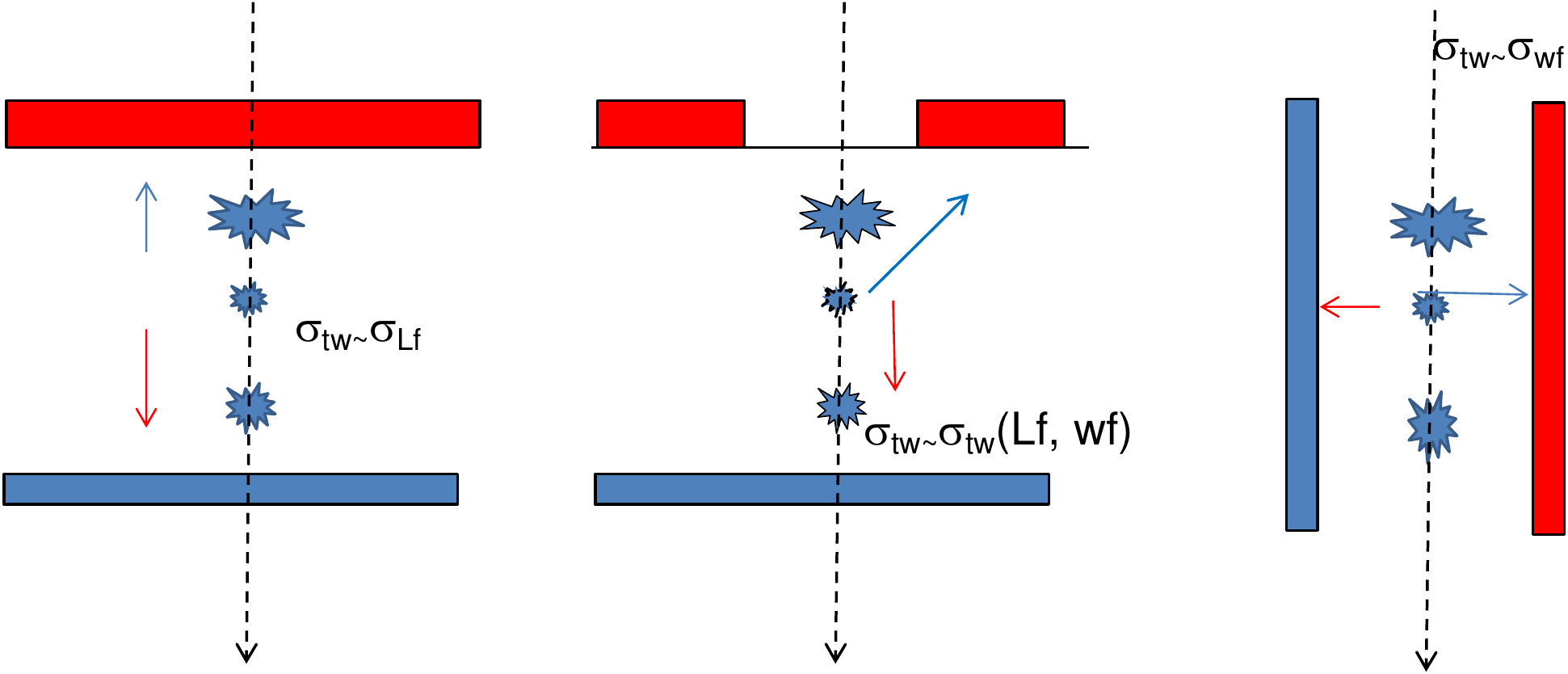,width=0.75\linewidth,clip=}\\
(a) \hspace{3.2cm} (b) \hspace{3.2cm} (c) \\
\end{center}
\caption{Illustration of the time walk contribution for different detectors types: (a) planar detector with thickness $<<$ cell size (b) finely segmented detector and (c) 3D detector.}
\label{fi:TimeWalkContributions}
\end{figure}

For planar detectors with thickness $\ll$ cell size the weighting field is constant over the entire cell 
and cannot cause any differences in signal shape (see Fig. \ref{fi:TimeWalkContributions}). Hence, the time walk is dominated by Landau 
fluctuations ($\sigma_{tw}\approx \sigma_{Lf}$), particularly for LGADs where electrons need to reach the gain layer to multiply. For fine segmentations, careful test beam studies 
can be used to separate both contributions, such as for the NA62 pixel detectors \cite{NA62}. In 
3D detectors Landau fluctuations are less important as charges generated at different depths have 
the same drift distance to the collection electrode. The time walk contribution 
is therefore dominated by the location of impact within the cell ($\sigma_{tw}\approx \sigma_{wf}$). 
This isn't entirely true for inclined tracks, but absence of gain and short drift distances render $\sigma_{Lf}$ to be negligible.

\section{Simulation of detectors}

\begin{figure}[!hbt]
\begin{tabular}{cc}
\epsfig{file=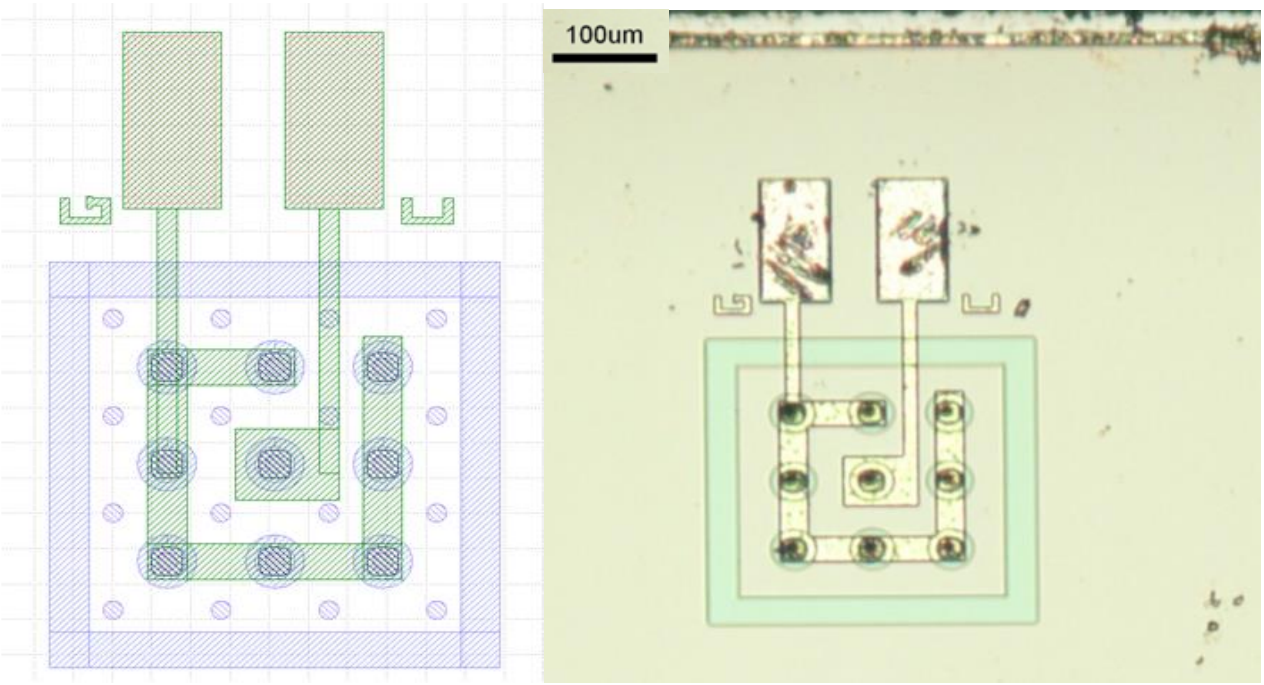,width=0.5\linewidth,clip=} & \epsfig{file=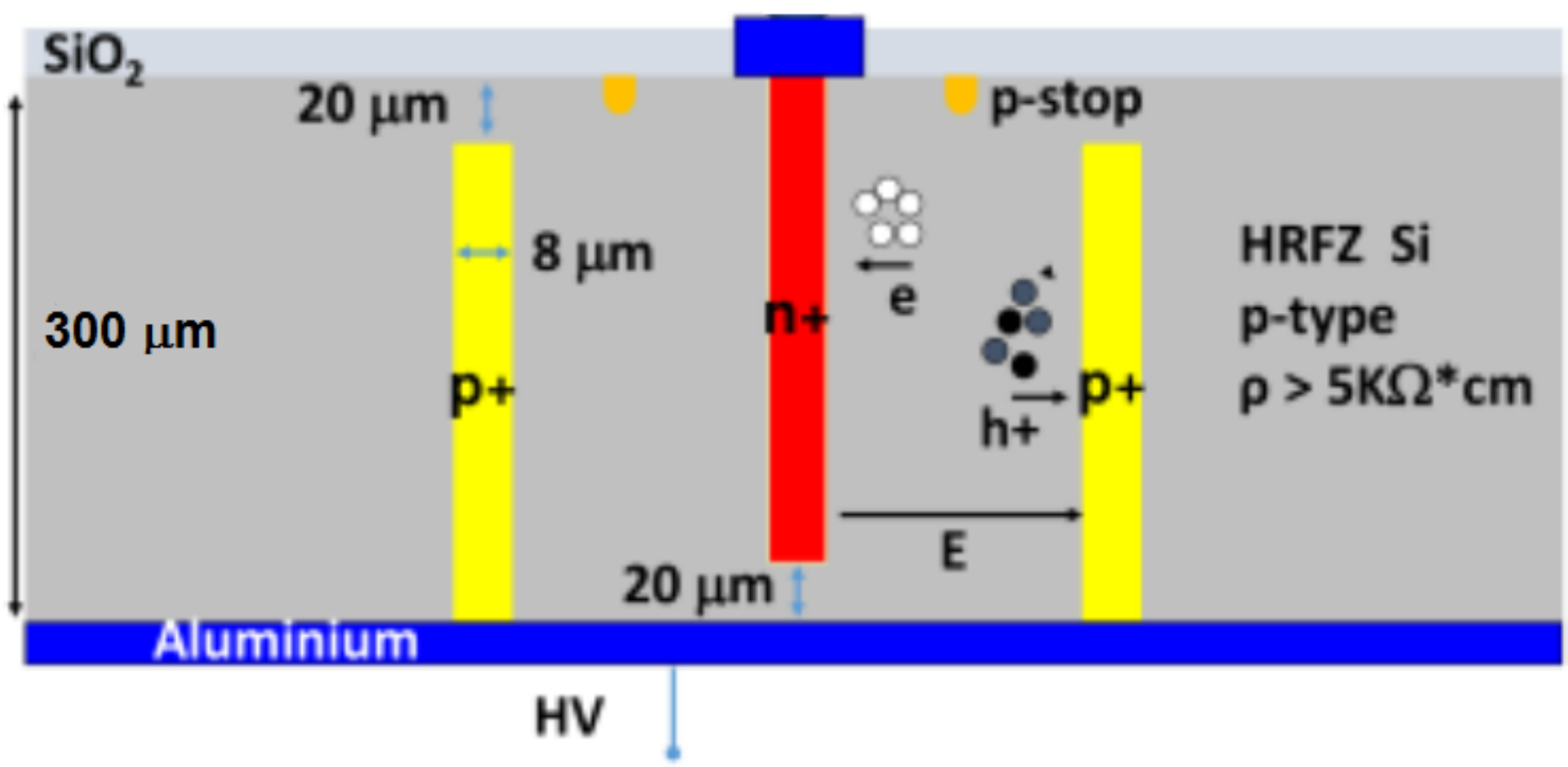,width=0.5\linewidth,clip=} \\
(a) & (b) 
\end{tabular}
\caption{(a) The design/photo of a single cell structure used in measurements and simulations. (b) The cross-section of 
the investigated 3D detector.}
\label{fi:Sample}
\end{figure}

A special structure produced by CNM was used in the studies and is shown in Fig. \ref{fi:Sample}a. 
A single 50x50 $\upmu$m cell with an $n^+$ readout electrode (1E) was surrounded by eight
neighboring cells connected together. The thickness of the investigated detector 
was 300 $\upmu$m with a $p$ type bulk resistivity 
of $\sim$5 k$\Omega$cm. The diameter of the holes was 8-10 $\upmu$m. The junction columns were etched from the top while 
the four ohmic columns at each corner of the cell were etched from the bottom. Both column types penetrate some 20 $\upmu$m short of the 
full thickness as shown in Fig. \ref{fi:Sample}b.

The software package KDetSim \cite{SimWWW} was used to simulate the 
charge collection in such 3D detector. The package solves the three dimensional 
Poison equation for a given effective doping concentration $N_{eff}$ to 
obtain the electric field and the Laplace equation for the weighting field. 
The induced current is calculated according to the Ramo's theorem \cite{Ramo39}, where 
the charge drift is simulated in steps with diffusion and trapping also taken into account. 
The minimum ionizing particle track was split into ``buckets'' of charge 1 $\upmu$m apart. 
The drift of each bucket was then simulated and the resulting 
induced current is the sum of all such contributions. The details of the simulation 
can be found in several references \cite{Sim1,Sim2}. The simulated induced 
current is then processed with a transfer function of a fast charge integrating preamplifier followed 
by a CR-RC$^3$ shaping circuit with a peaking time of 1 ns. 
\begin{figure}[!hbt]
\begin{tabular}{cc}
\epsfig{file=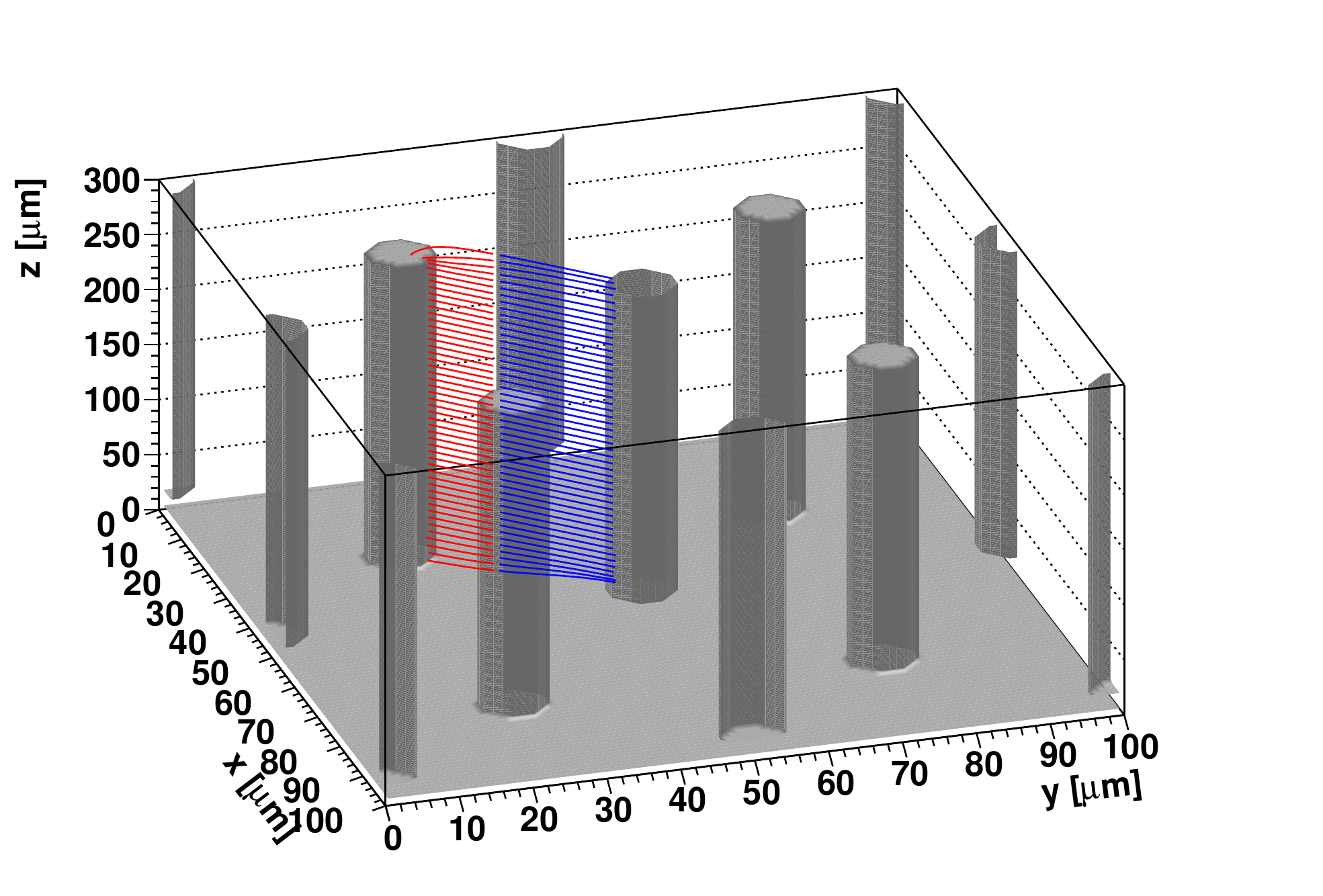,width=0.52\linewidth,clip=} & \epsfig{file=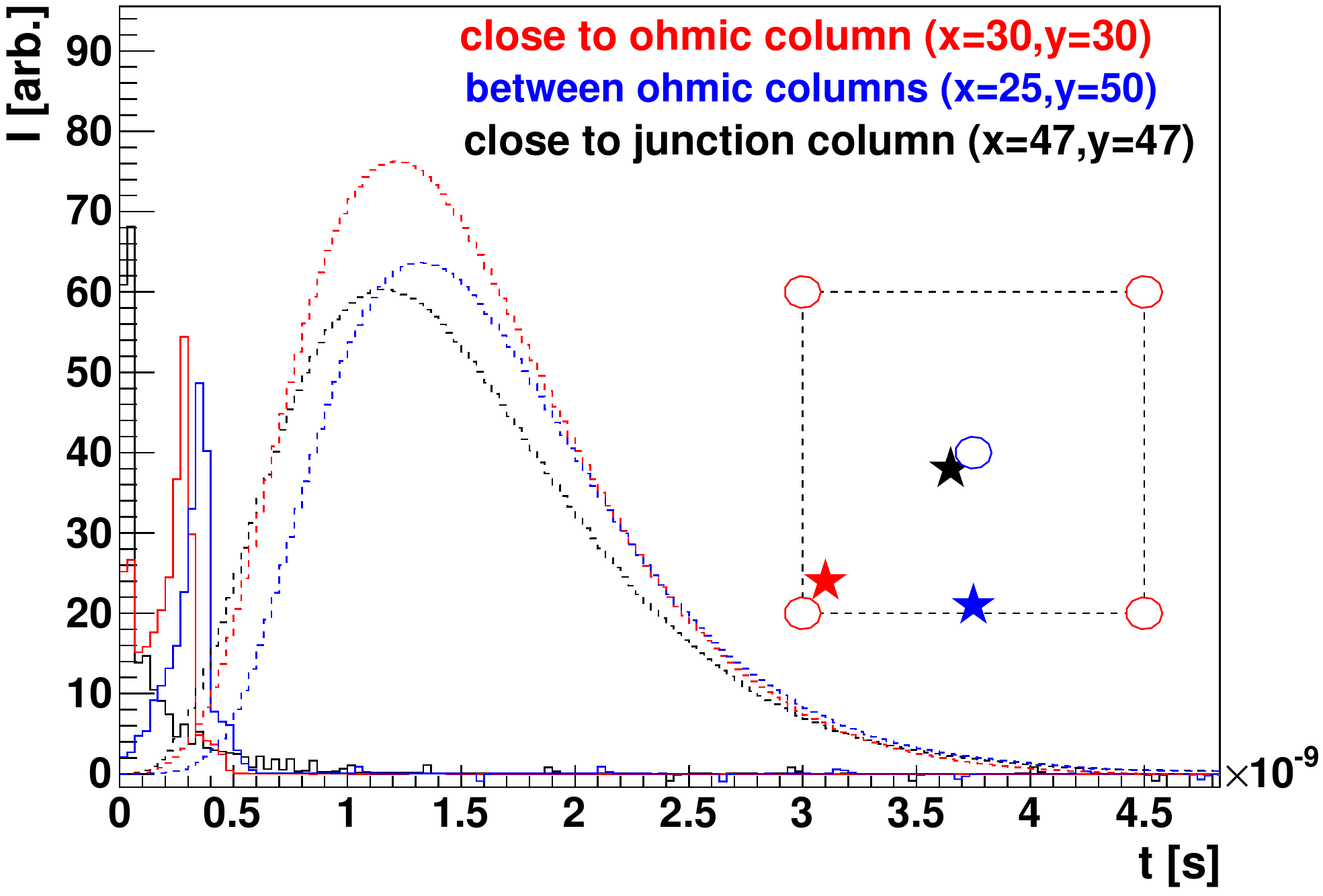,width=0.48\linewidth,clip=} \\
(a) & (b) 
\end{tabular}
\caption{(a) Simulated drift paths of electrons (blue) and holes (red) for a perpendicular track. (b) Currents induced for perpendicular hits at different positions (solid lines) and signals after electronics response (dashed lines) at 50 V.}
\label{fi:TrackSim}
\end{figure} 

An example of a minimum ionizing particle hitting the cell is shown 
in Fig. \ref{fi:TrackSim}a. The drift paths of electrons (blue) to 
junction $n^+$ column and holes (red) to ohmic $p^+$ columns are 
shown. The obtained induced currents for three different hit 
positions indicated in the picture (solid lines) and the signal after 
electronics processing (dashed lines) are shown in Fig. \ref{fi:TrackSim}b. 
In the simulation the constant fraction discrimination with 25\% fraction 
was used to determine the time of arrival (ToA). A charge of at least 1000 $e_0$ was required 
to actually calculate the time stamp of the hit (the hits with less have ToA=0). 
Several different amplification circuit models (CR-RC$^n$) were used with 
different peaking times which all yield similar results.  

\begin{figure}[!hbt]
\begin{tabular}{cc}
\epsfig{file=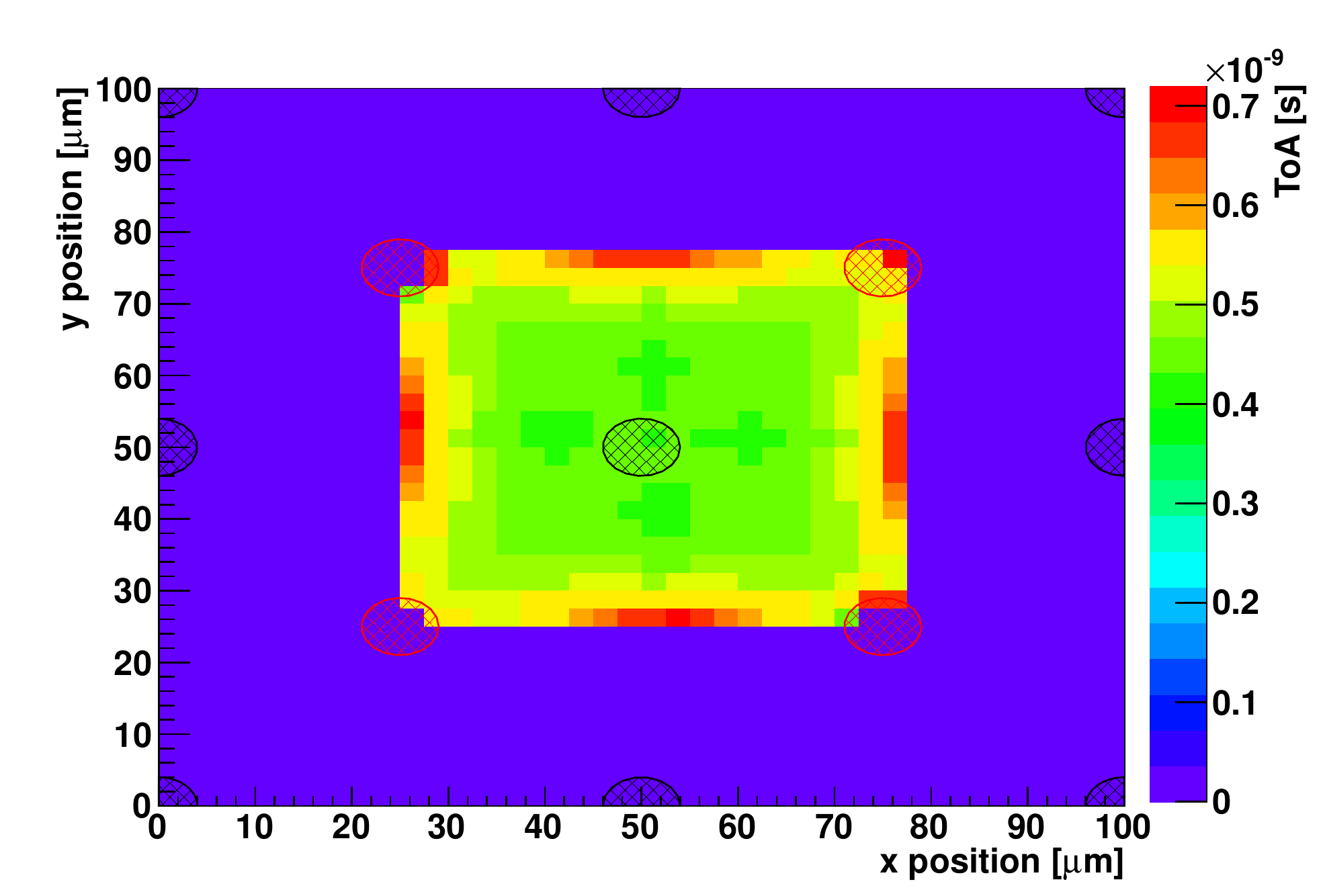,width=0.5\linewidth,clip=} & \epsfig{file=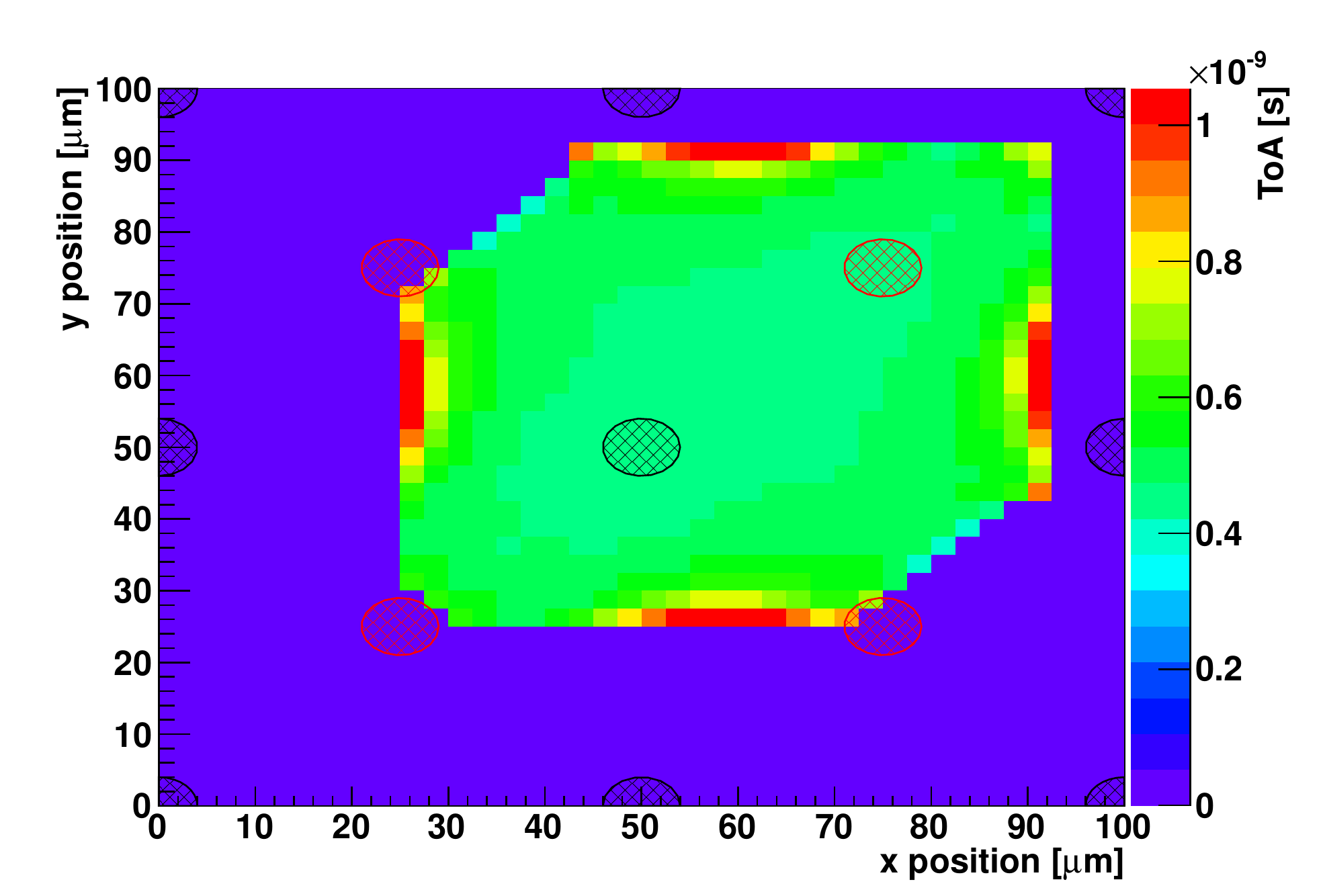,width=0.5\linewidth,clip=} \\
(a) & (b) \\
\epsfig{file=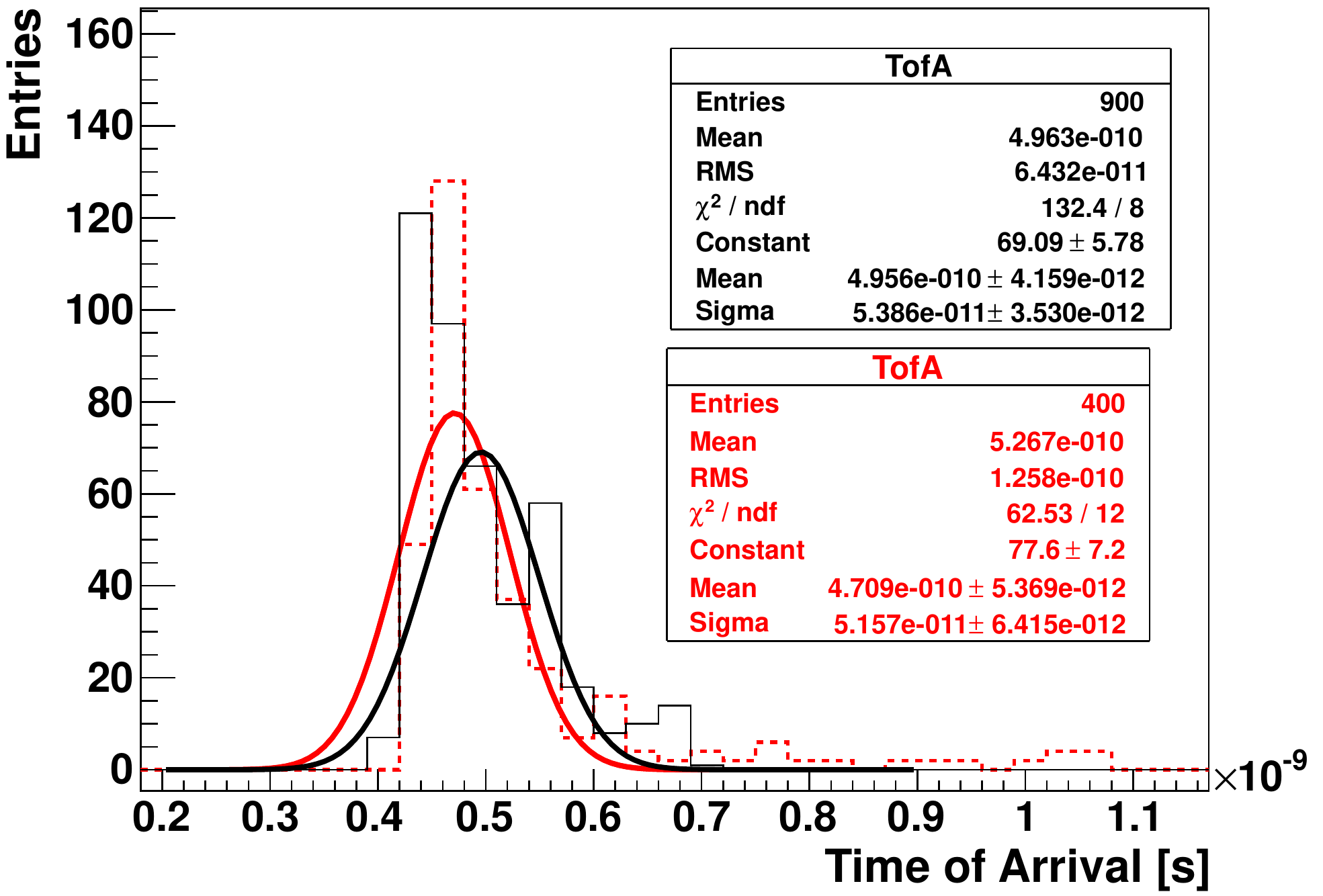,width=0.5\linewidth,clip=} & \epsfig{file=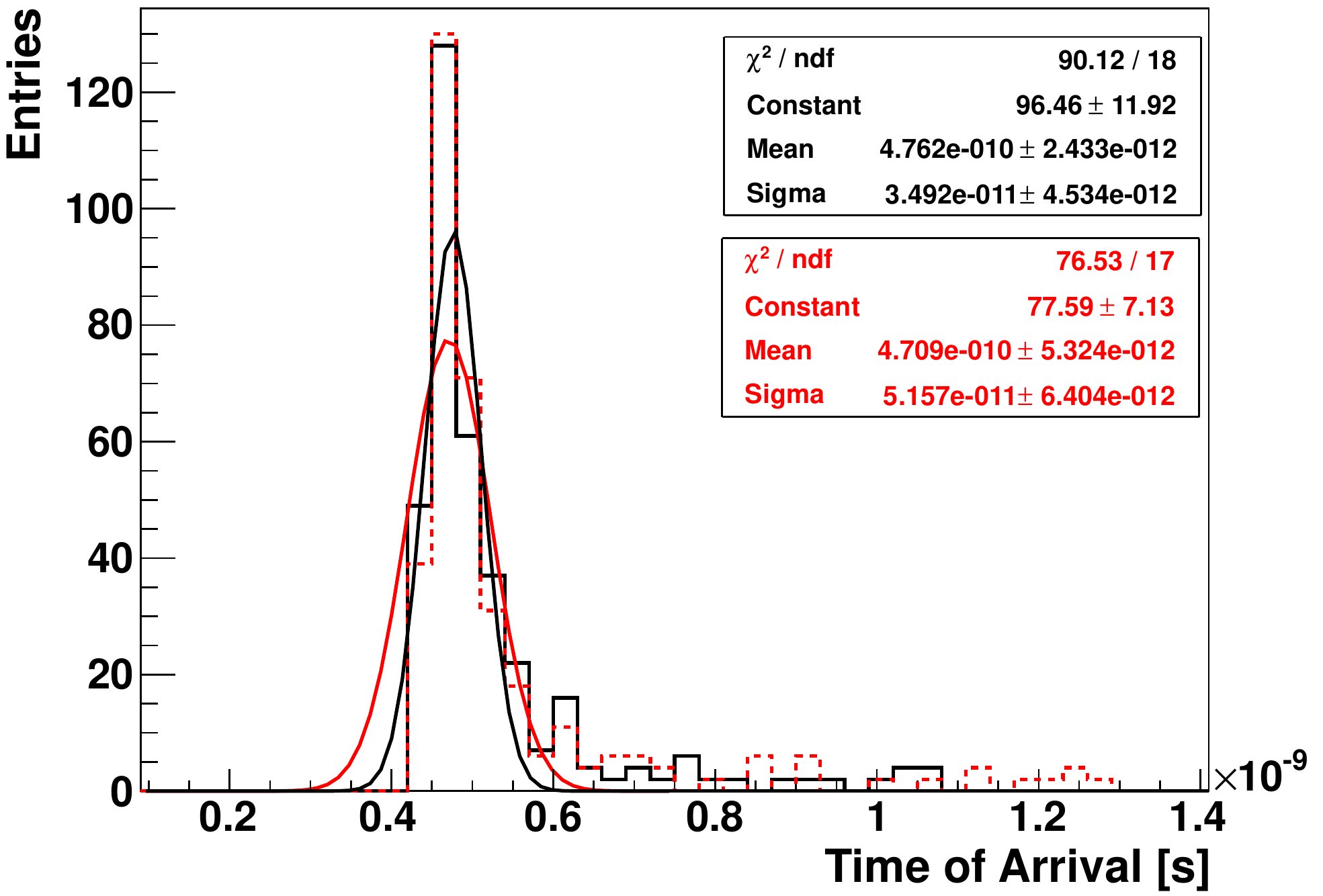,width=0.5\linewidth,clip=} \\
(c) & (d) 
\end{tabular}
\caption{Time of arrival for different hit positions at 50 V, 27$^\circ$C for: (a) perpendicular tracks and (b) tracks under small angle, 5$^\circ$ (equal inclination in $x$ and $y$). (c) Distribution of the ToA and Gaussian fit to it for perpendicular tracks (black) and  tracks under 5$^\circ$ angle (red). (d) Same as (c) for inclined tracks at $T=27^\circ$C (dashed red) and $T=-20^\circ$C (solid black).}
\label{fi:TofA2D}
\end{figure}

ToA for perpendicular tracks with impact positions distributed across the cell 
are shown in Fig. \ref{fi:TofA2D}a. 
As expected the signal for tracks hitting the regions with a saddle in 
the electric field (between ohmic and junction electrodes)
showed a delayed ToA. En example of ToA map for ionizing particles 
under $5^\circ$ angle is shown in Fig. \ref{fi:TofA2D}b.  

The histogram of ToA over the cell surface for both cases is shown 
in Fig. \ref{fi:TofA2D}c. The width of the Gaussian fit to the peak 
of the distribution is an estimate of the hit-position contribution 
to the time resolution ($\sigma_{wf}$). The distribution is not symmetrical 
and has a tail which is larger for a larger angle, although the Gaussian width is 
smaller. The hit position contribution is $\sigma_{wf} \sim 54$ ps for perpendicular 
tracks and 51 ps for tracks under an angle of 5$^\circ$ at 50 V and room temperature. Decrease 
of the temperature improves the time resolution substantially as shown in 
Fig. \ref{fi:TofA2D}d, due to a faster drift.

\begin{figure}[!hbt]
\begin{tabular}{cc}
\epsfig{file=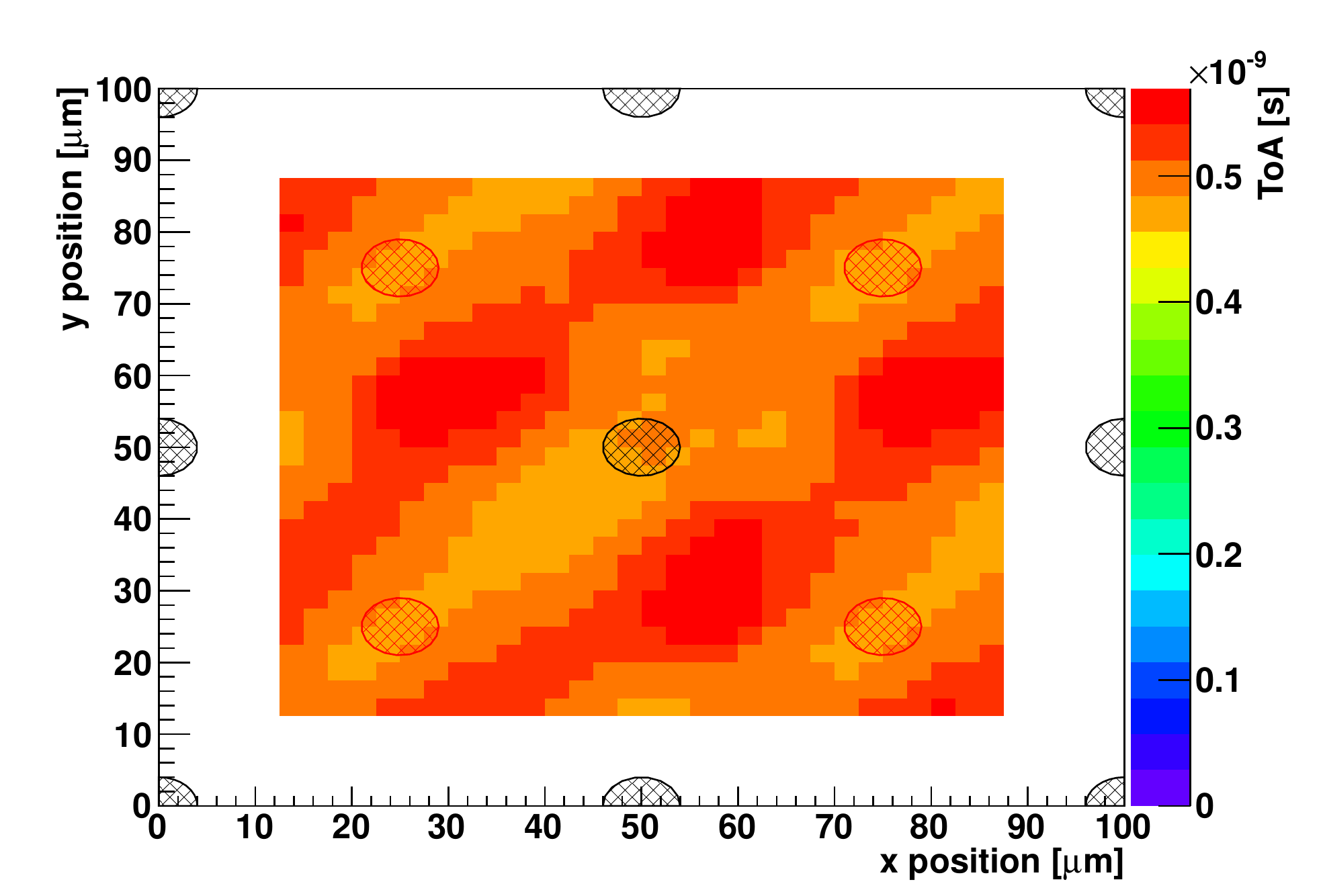,width=0.5\linewidth,clip=} & \epsfig{file=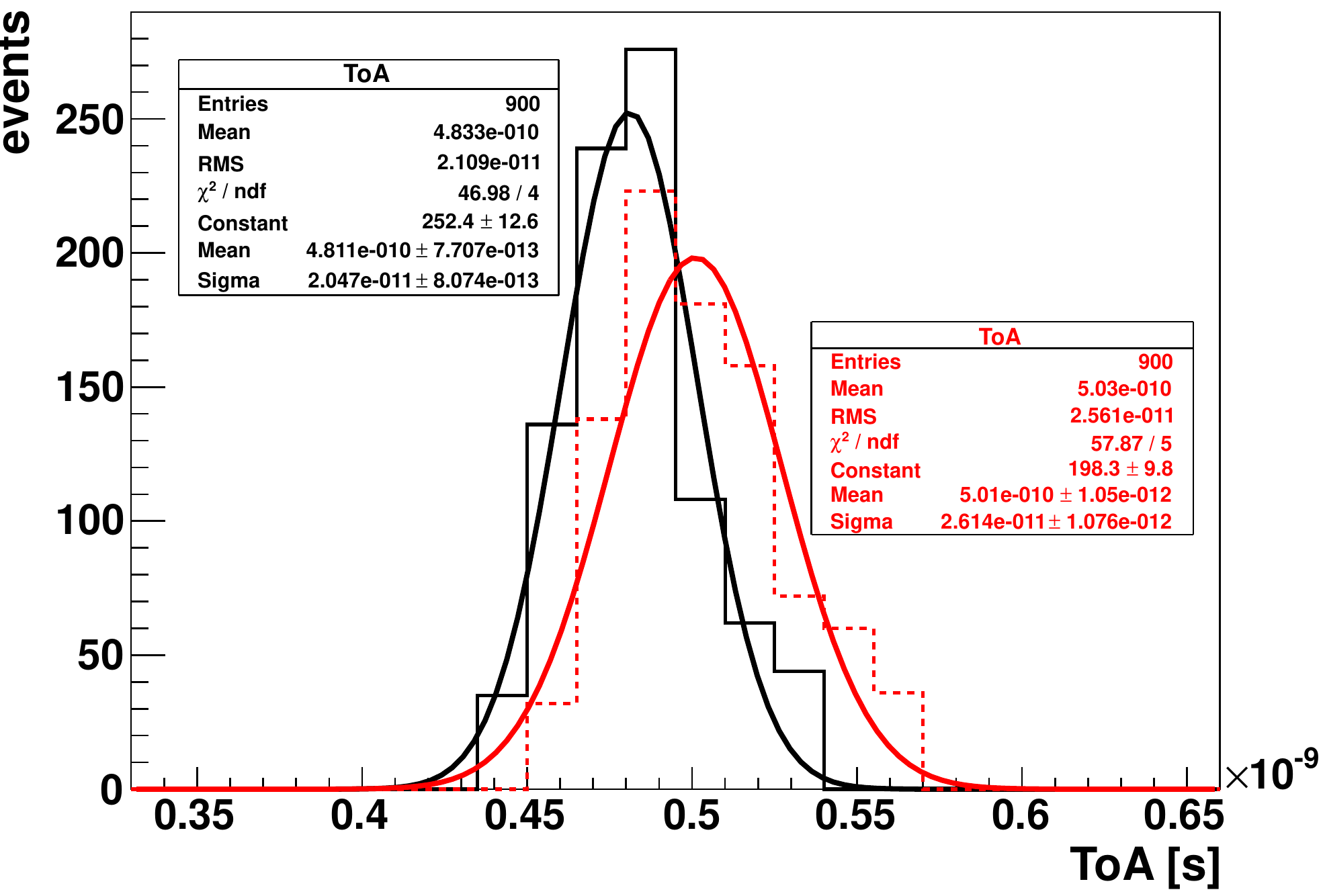,width=0.5\linewidth,clip=} \\
(a) & (b) 
\end{tabular}
\caption{(a) ToA for different inclined tracks hit positions at 50 V, 27$^\circ$C, where all the cells are connected together. (b) Distribution of ToA and Gaussian fit to it for the inclined tracks at $T=27^\circ$C (dashed red) and at $T=-20^\circ$C (solid black).}
\label{fi:TofA2DWFALL}
\end{figure}

The distributions in Figs. \ref{fi:TofA2D} refer to the case where the cells are read out 
separately. If multiple cells are connected together then the charge sharing, which 
reduces a single cell signal, does not occur and a much narrower 
distribution is obtained without tails as can be seen in 
Fig. \ref{fi:TofA2DWFALL}. Around $\sigma_{wf}\sim 26$ ps is obtained for inclined tracks ($5^\circ$) at 50 V already at room temperature 
and $\sigma_{wf} \sim 20$ ps at -20$^\circ$C.

The simulation was used to predict the time resolution limits ($\sigma_{j} \ll \sigma_{wf}$) for different cell sizes, 
temperatures and doping concentrations. 
In this study separately readout square cells with a single junction column (1E) were assumed. 
The simulation was done for perpendicular tracks at a temperature of -20$^\circ$C. Instead of the width from a  
Gaussian fit a more conservative estimation, RMS of the ToA distribution was used 
as  $\sigma_{wf}$.  

\begin{figure}[!hbt]
\begin{tabular}{cc}
\epsfig{file=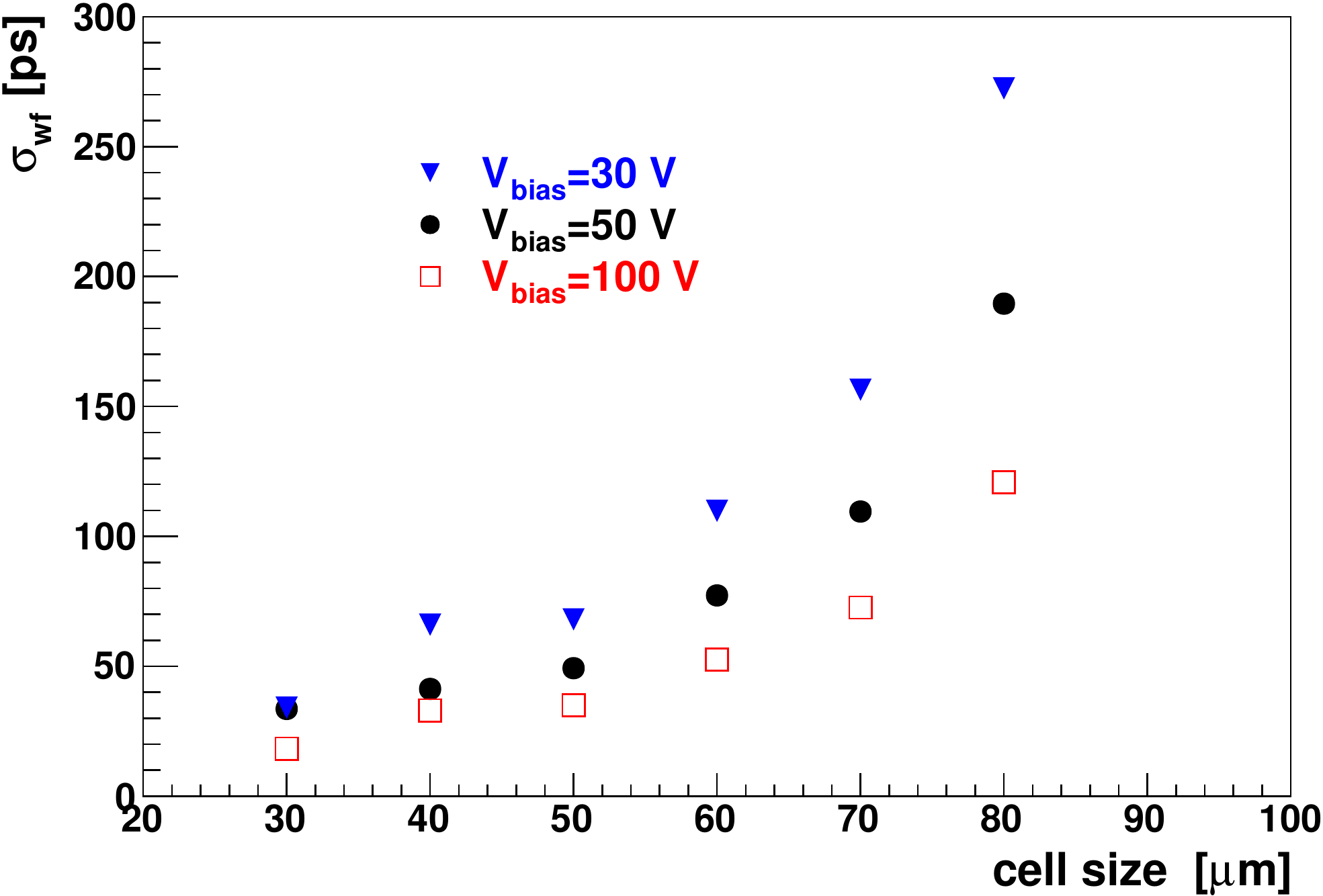,width=0.5\linewidth,clip=} & \epsfig{file=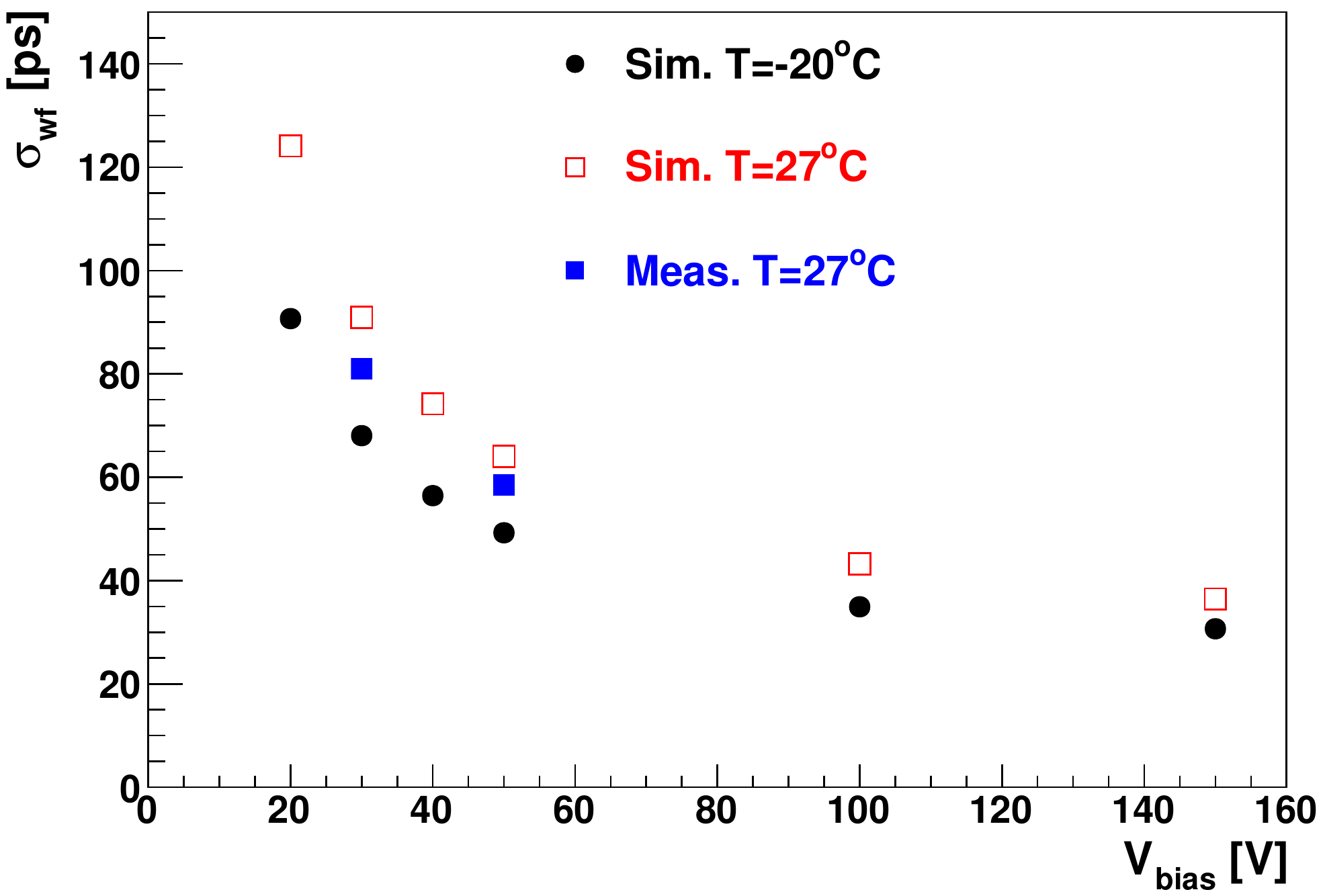,width=0.5\linewidth,clip=} \\
(a) & (b)
\end{tabular}
\begin{center}
\epsfig{file=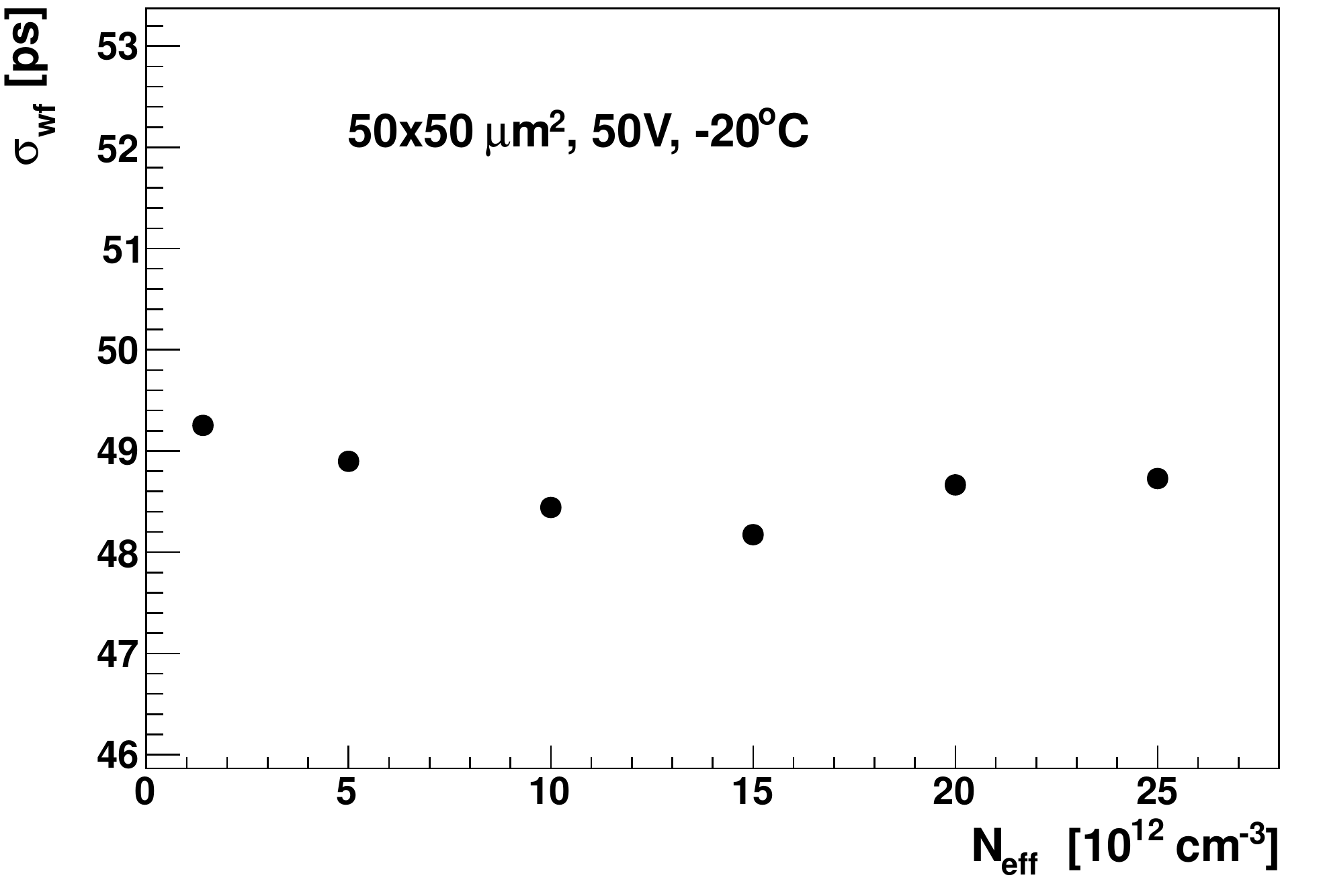,width=0.5\linewidth,clip=} \\
(c)
\end{center}
\caption{Dependence of $\sigma_{wf}$ on: (a) square cell length for different bias voltages at -20$^\circ$C, (b) bias voltage for a $50\times50$ $\upmu$m$^2$ cell at different temperatures, (c) effective doping concentration of the wafer. Measured points are also shown with solid squares in (b).}
\label{fi:CellSizeAndBias}
\end{figure}

The dependence of $\sigma_{wf}$ on cell size is shown in 
Fig. \ref{fi:CellSizeAndBias}a for different bias voltages. For large cell sizes 
the time resolution degrades rapidly, particularly at small bias voltages. The 
faster drift time at lower temperatures improves the time resolution for 
a $50\times50\,\,\upmu$m$^2$ cell as shown in Fig. \ref{fi:CellSizeAndBias}b. 
At 50 V  $\sigma_{wf}$ ranges from 46 ps at -20$^\circ$C to 63 ps at 27$^\circ$C. At 
even higher bias voltages of 100 V $\sigma_{wf} \sim 35$ ps is achieved. Almost no 
influence of wafer doping concentration on $\sigma_{wf}$ is predicted by simulations as shown 
in Fig. \ref{fi:CellSizeAndBias}c.

\section{Measurements}

\begin{figure}[!hbt]
\begin{center}
\epsfig{file=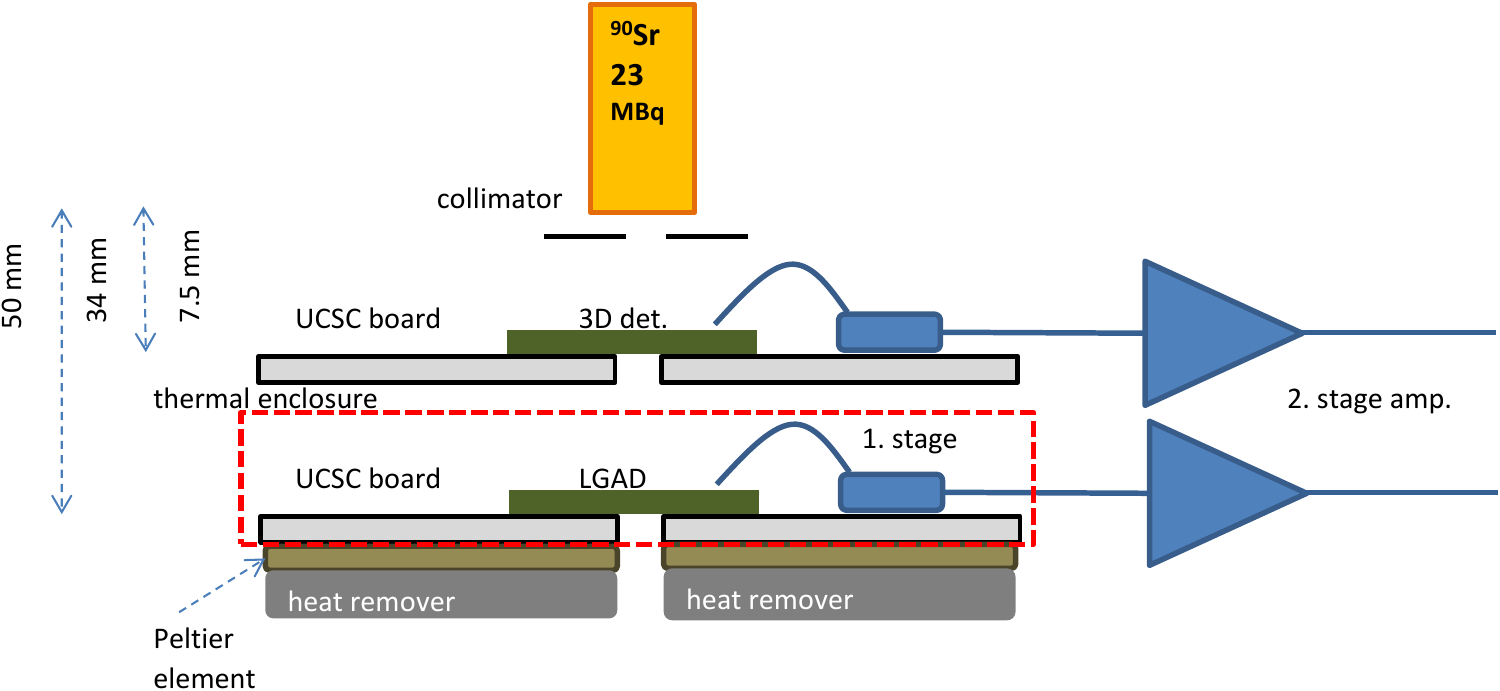,width=0.75\linewidth,clip=}\\
\end{center}
\caption{Experimental setup used for measurements of the time 
resolution of the $50 \times 50$ $\upmu$m$^2$ 3D detector.}
\label{fi:Setup}
\end{figure}

The experimental setup is shown in Fig. \ref{fi:Setup}. 
Electrons from $^{90}$Sr source ($E_{max}$=2.3 MeV) were used to determine 
the time resolution of the test-structure shown in Fig. \ref{fi:Sample}a. 
The central columns ($n^+$)of two such test-structures 400 $\upmu$m apart were 
connected together to the input of the amplifier, while the neighboring 
$n^+$ electrodes were grounded. The first stage of amplification uses fast trans-impedance 
amplifier designed by UCSC \cite{UCSC} followed by a second amplifier 
which gives signals large enough to be relatively easily recorded by a 
40 GS/s digitizing oscilloscope with 2.5 GHz bandwidth. 

The reference time required for measurement of the time resolution was 
provided from a non-irradiated LGAD detector produced by HPK \cite{GK2018,UCSC}. 
It is $50\,\,\upmu$m thick, has a diameter of 0.8 mm and high gain 
of $\sim 60$ at 330 V and room temperature. The time resolution of HPK-50D sensors 
was measured (in the same setup with a method described below) by using two such 
detectors and was determined to be 26 ps. More details on timing 
and charge collection measurements with these LGADs can be found in 
\cite{GK2018,UCSC}. The electrons were collimated to an angle of $<1^\circ$ 
by a combination of collimator, small cell size and the circular opening in the 
PCB boards hosting the sensors.

The trigger was provided by the reference LGAD sensor, but due to small surface of the 
3D detector ($2 \times 50 \times 50\,\,\upmu$m$^2$) the signal equivalent to half of the 
most probable signal, corresponding to five times the noise, was required in the 3D detector as well. 
Without that requirement the majority of triggers would be without hits in 
the 3D detector. The measurements were done at room temperature for both sensors. The 
3D detectors used in the measurements had a breakdown votlage of slightly more than 50 V 
which  prevented studies at higher voltages. 

\begin{figure}[hbt]
\begin{tabular}{cc}
\epsfig{file=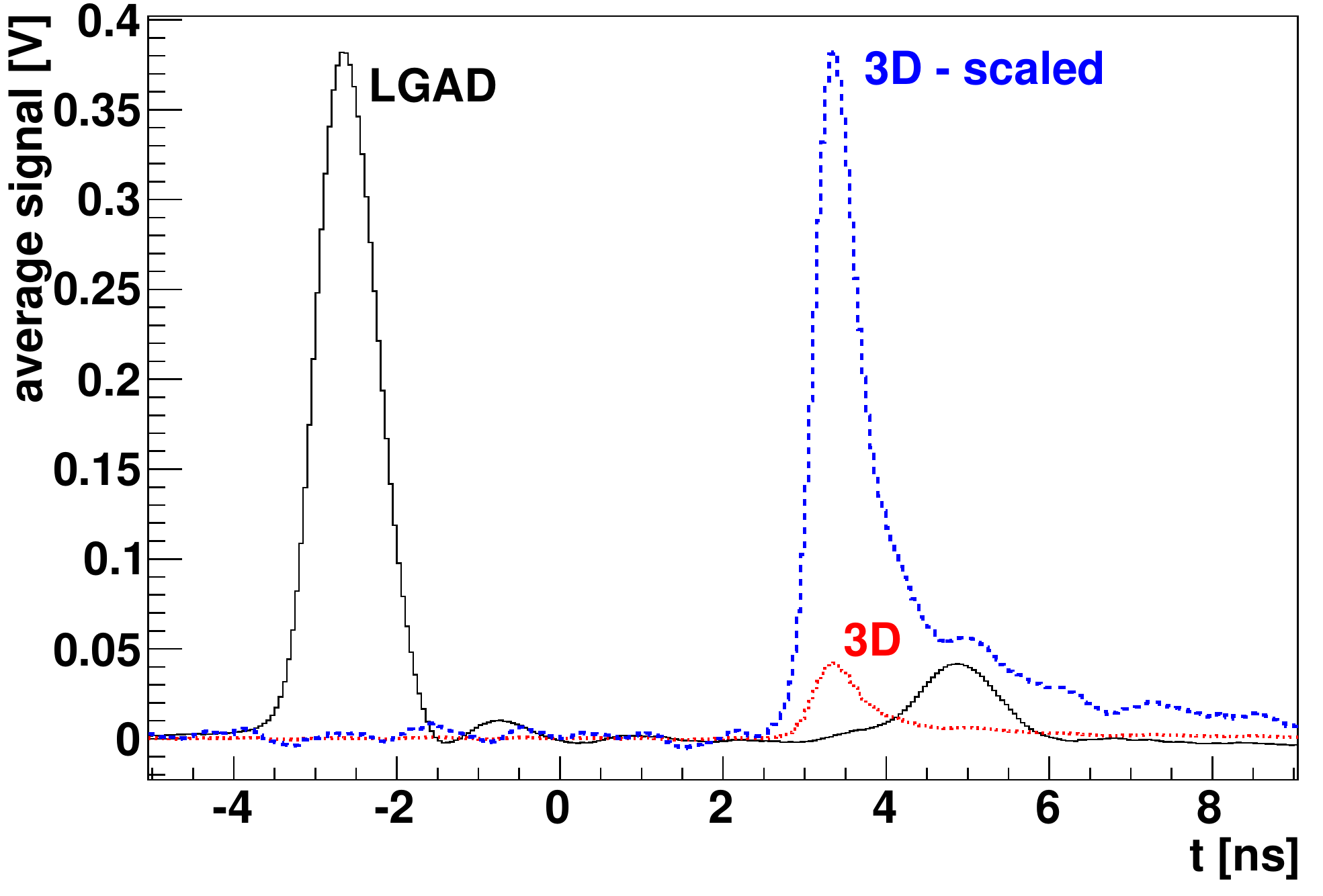,width=0.5\linewidth,clip=} & \epsfig{file=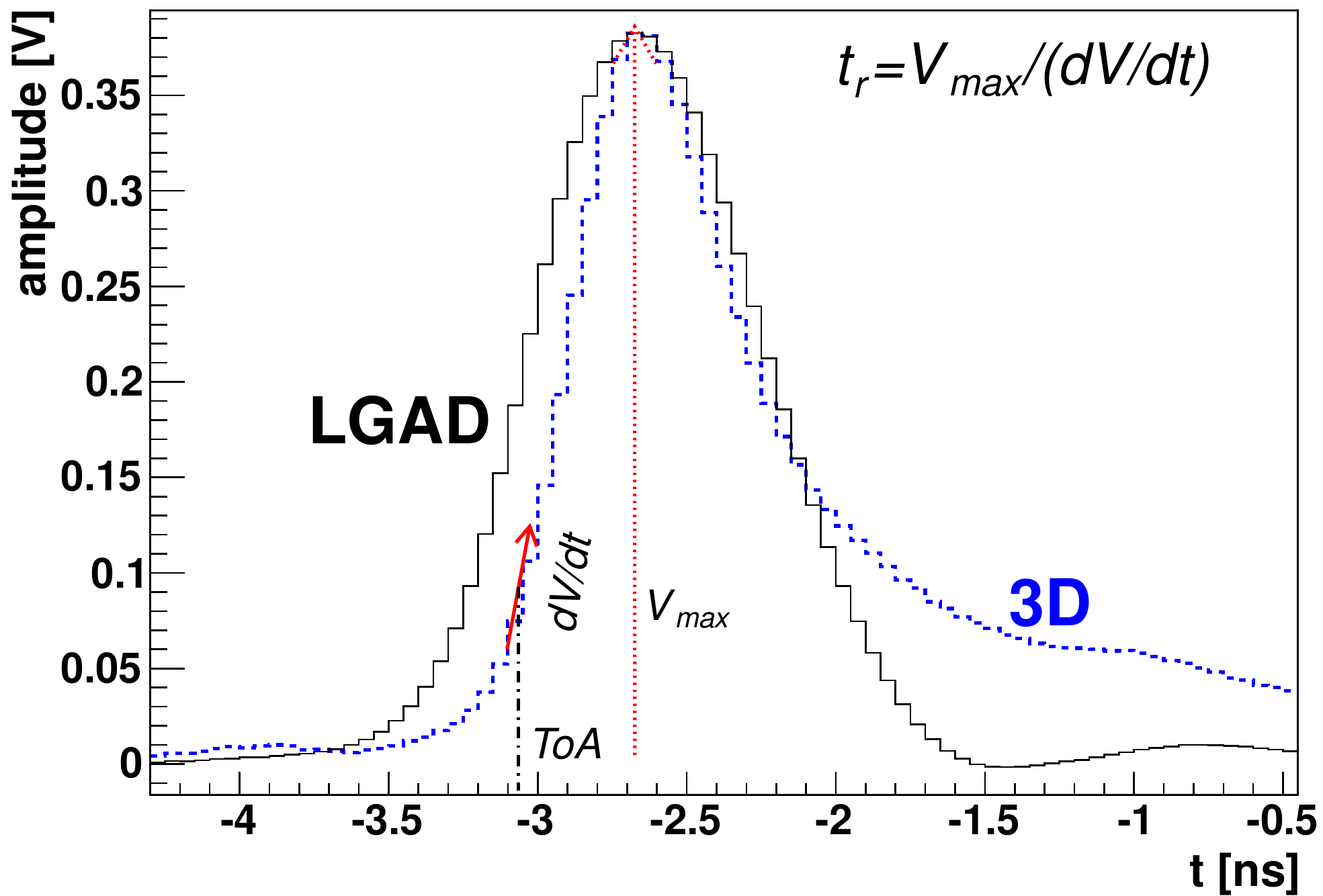,width=0.5\linewidth,clip=} \\
(a) & (b) \\
\epsfig{file=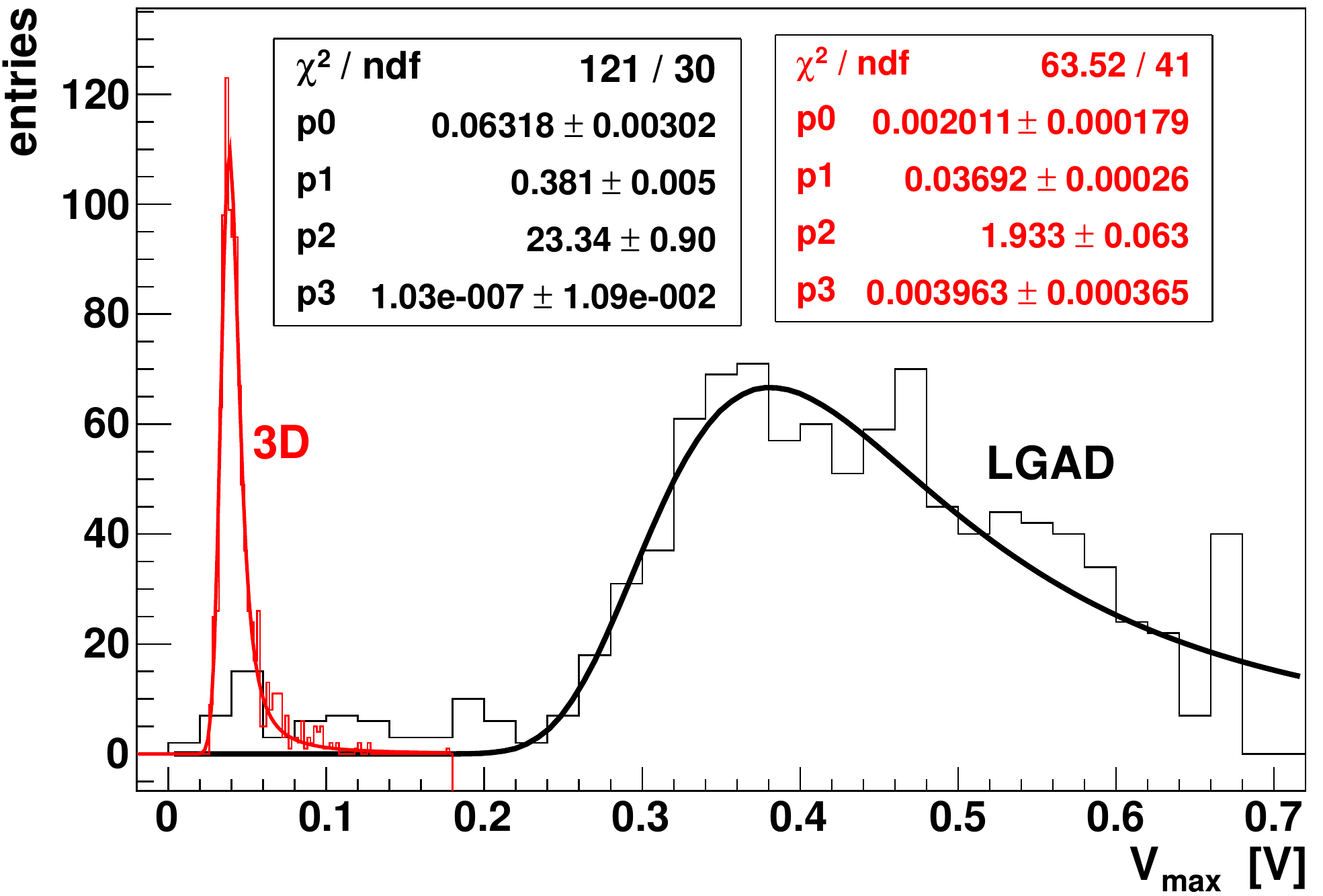,width=0.5\linewidth,clip=} & \epsfig{file=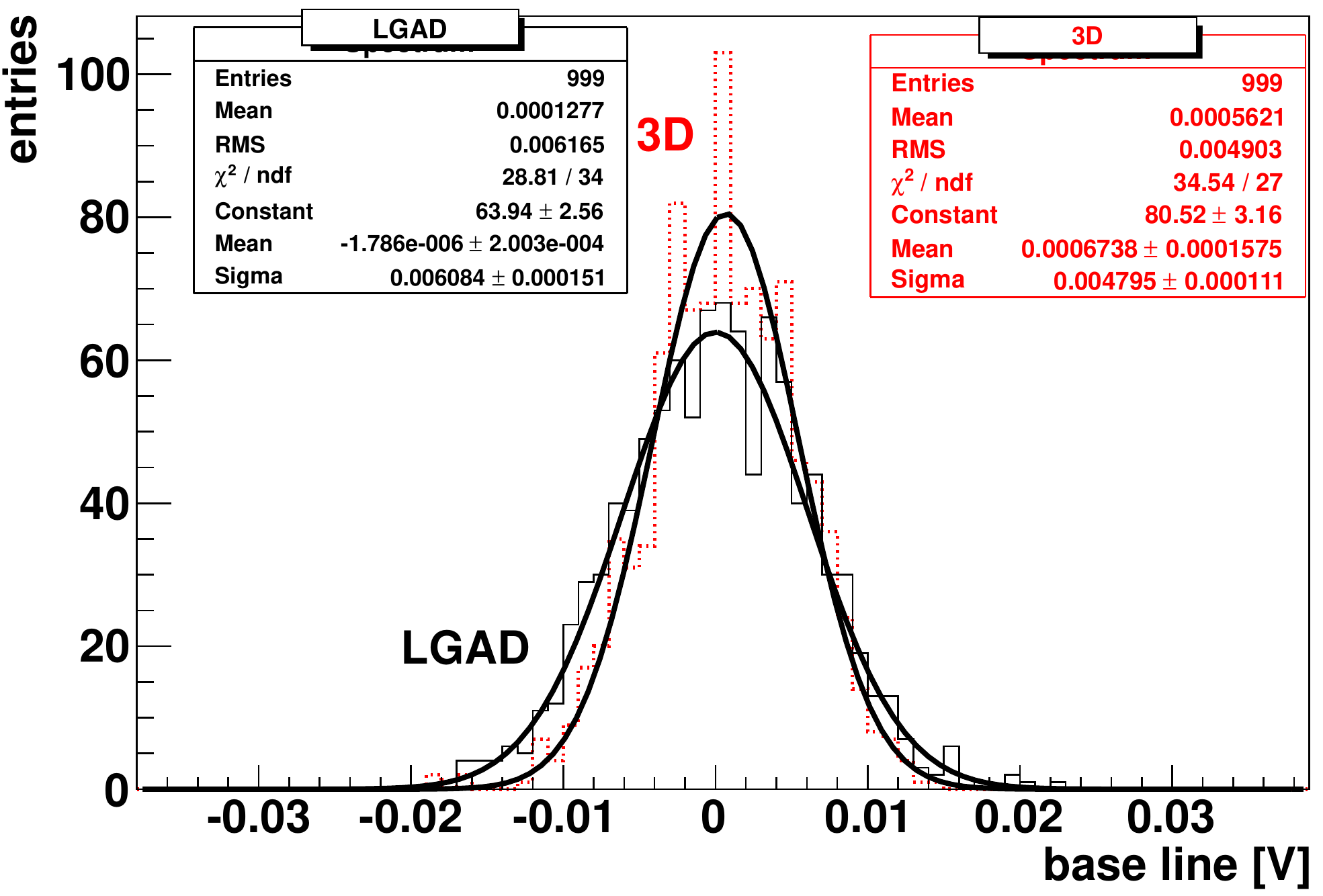,width=0.5\linewidth,clip=} \\
(c) & (d) \\
\end{tabular}
\caption{Comparison of 3D and LGAD detectors : (a) average of recorded waveforms, (b) 3D waveform scaled and aligned with LGAD peak; note the indication of quantities used in text, (c) signal spectrum and (d) noise distribution. 3D detector was biased to 50 V.}
\label{fi:PulseShapes}
\end{figure}

The comparison of averaged pulses from LGAD and 3D detector is shown 
in Fig. \ref{fi:PulseShapes}a. Apart from the obvious difference in height 
due to a large gain of the LGAD, the difference 
in both pulse shapes can be noticed. The slew rate is steeper for the 3D detector 
(see Fig. \ref{fi:PulseShapes}b), which however 
exhibits a longer tail. 

The spectrum of deposited charge, measured as amplitude of the signal $V_{max}$, for the  
LGAD and the 3D detector are compared in Fig. \ref{fi:PulseShapes}c. The fit of the convoluted Landau and 
Gaussian distributions to the data is also shown. The difference of factor $\sim 10$ 
was observed in the most probable signal (parameter $p1$ of the fit) which is also expected from 
the measured gain \cite{GK2018} and the difference in the detector thickness. As the noise level 
is 20\% larger for LGAD, due to higher capacitance (see Fig. \ref{fi:PulseShapes}d), the 
difference in $S/N$ is by a factor of $\sim 8.5$. 

\begin{figure}[hbt]
\begin{tabular}{cc}
\epsfig{file=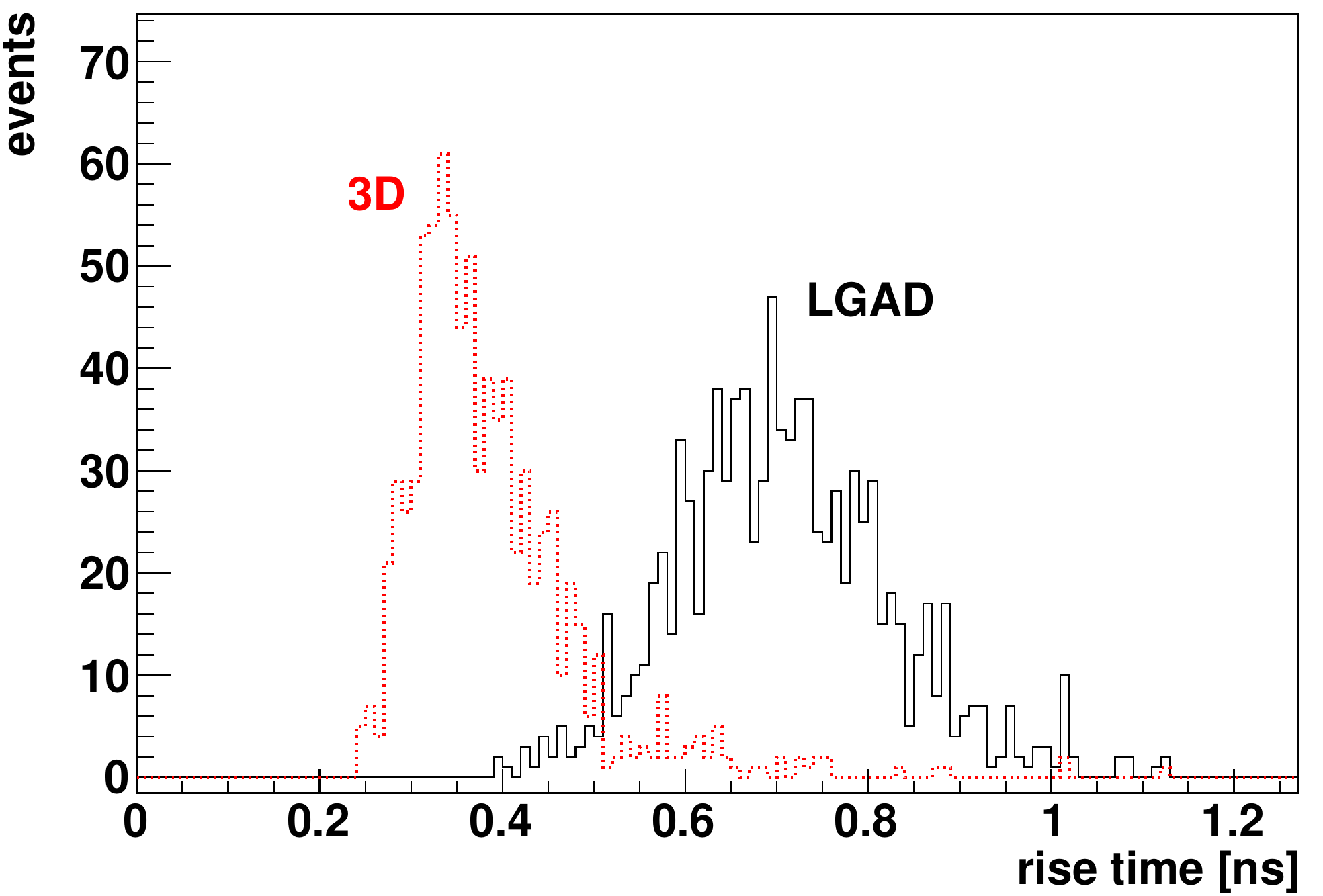,width=0.5\linewidth,clip=} & \epsfig{file=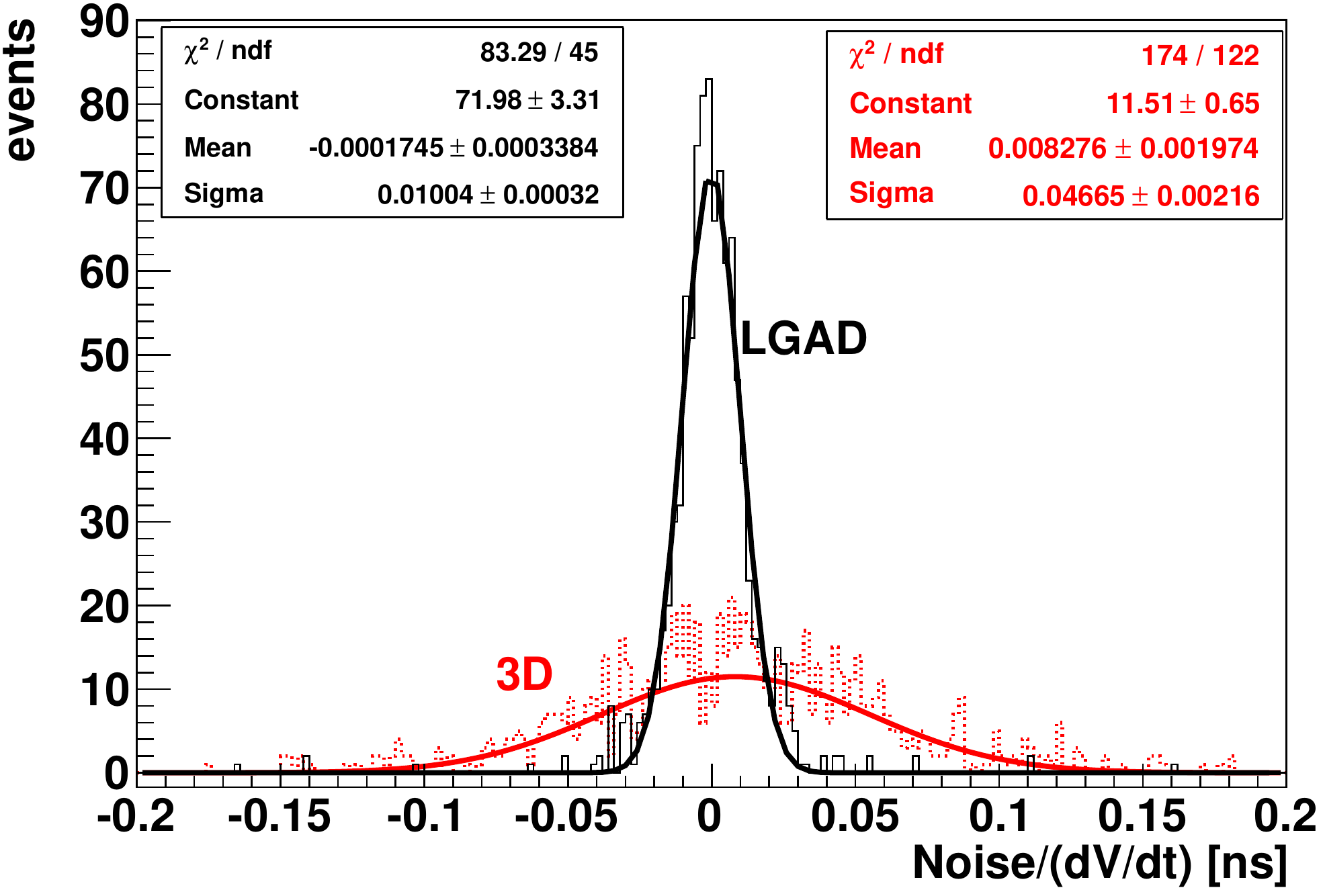,width=0.5\linewidth,clip=} \\
(a) & (b) \\
\end{tabular}
\caption{(a) Comparison of the rise time in LGAD and 3D detector and (b) measured jitter - $\sigma_j$.}
\label{fi:Jitter}
\end{figure}

Faster rise time of the signal for the 3D detector is beneficial as it leads to a smaller jitter. 
For each event the rising edge was fitted with linear function around ToA ($dV/dt$) and the rise time was  
determined as $t_{rise}=V_{max}/(dV/dt)$ (see Fig. \ref{fi:PulseShapes}b).  The distribution 
of the rise times is shown in Fig. \ref{fi:Jitter}a. Such 3D 
detectors have therefore the rise time around two times shorter than 50 $\upmu$m thick LGADs. 
This is also reflected in the jitter measurement shown in Fig. \ref{fi:Jitter}b. The measured jitter 
for LGAD detectors ($\sigma_{j,LGAD}=10$ ps) is 4.7 times smaller than that of the 3D detector ($\sigma_{j,3D}=47$ ps). 
This is in agreement with Eq. \ref{eq:TimeRes} when the difference in 
average rise time by a factor of 1.75 and in $S/N$ by a factor of 8.5 is used.

\begin{figure}[hbt]
\begin{center}
\epsfig{file=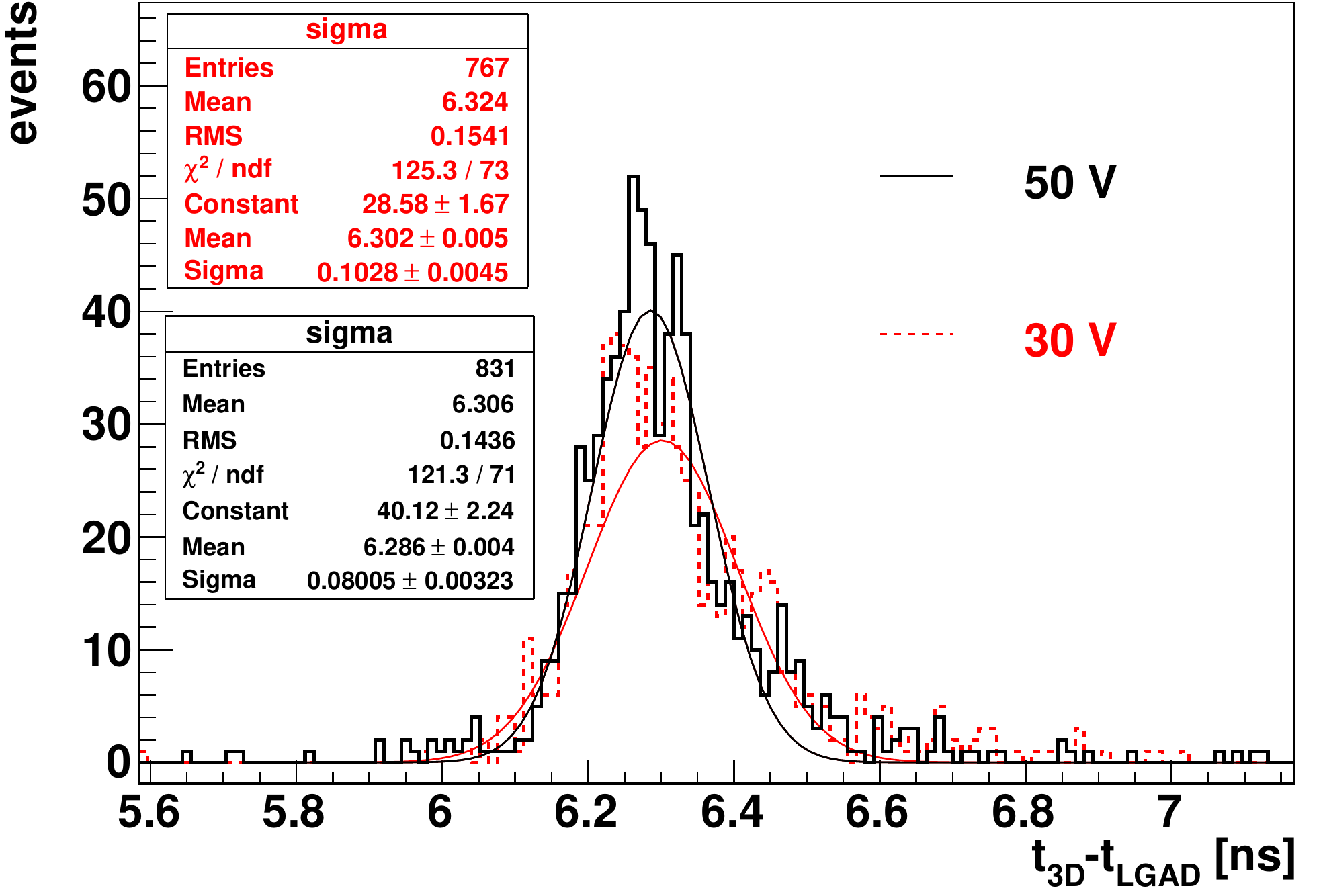,width=0.65\linewidth,clip=}  
\end{center}
\caption{Measured ToA difference between LGAD and 3D detector ($T=27^\circ$C). }
\label{fi:TimeRes}
\end{figure}

The time resolution was measured as the difference in ToA between LGAD 
(reference detector) and the 3D detector. As in simulation, ToA was measured 
when 25\% of the maximum signal was reached (CFD). The distribution of the time 
difference $t_{3D}-t_{LGAD}$ is  shown in Fig. \ref{fi:TimeRes} for 
30 V and 50 V. It was fit with Gaussian function and the extracted $\sigma_t$ was used 
as the time resolution of the measurement. It is given by
\begin{equation}
\sigma_{t}^2=\sigma_{LGAD}^2+\sigma_{3D}^2 \quad .
\label{eq:sigma_3D}
\end{equation}
The time resolution of LGAD detector is $\sigma_{LGAD}=26$ ps 
($\sigma_{j,LGAD}=10$ ps, $\sigma_{Lf}=25$ ps) therefore $\sigma_t$ 
is dominated by the time resolution of the 3D detector ($\sigma_{3D}$) with 
$\sigma_{3D}(50\,\,\mathrm{V})=75\,\, \mathrm{ps}$ and $\sigma_{3D}(30\,\,\mathrm{V}) =98\,\, \mathrm{ps}$.
With known $\sigma_{3D}$, $\sigma_{wf}$ can be calculated as
\begin{equation}
\sigma_{wf}^2 \approx \sigma_{3D}^2-\sigma_{j,3D}^2 \quad ,
\label{eq:sigma_wf}
\end{equation}
which gives $\sigma_{wf}(50\,\,\mathrm{V}) \approx 58 \,\, \mathrm{ps}$ and 
 $\sigma_{wf}(30\,\,\mathrm{V}) \approx 81 \,\, \mathrm{ps}$. This agrees within 10\% with the simulated 
values of 54 ps and 89 ps shown in Figs. \ref{fi:TofA2D}c and \ref{fi:CellSizeAndBias}b. 
Moreover, the tail in the timing distribution predicted in simulations is also observed in 
the measurements.

\section{Discussion}
A good agreement between measurements and simulations can be used to predict 
the operation also after irradiations. Charge 
collection measurements in similar pixel detectors showed a degradation by only 
a few percent at 175 V for 230 $\upmu$m thick detectors 
irradiated to an equivalent fluence of $\Phi_{eq}=1 \cdot 10^{16}$ cm$^{-2}$ \cite{RD50-CNM3D-1}. 
Also a superb detection efficiency of $>98\%$ was measured in test beam even after 
$\Phi_{eq}=2.8 \cdot 10^{16}$ cm$^{-2}$ at very high bias voltages of 200 V \cite{JLange2018}.  

The increase of the breakdown voltage with irradiation will improve the timing performance. 
If at the same time charge collection degradation is small and sufficient cooling 
is provided to keep also the leakage current and shot noise small, the timing resolution of 
irradiated detectors may even surpass that of the non-irradiated one. Even more so, if 
3D detectors will be operated in charge multiplication mode \cite{RD50-CNM3D-1}.

At the same time 100\% fill factor can be maintained with 3D detectors. If larger 
cells are used the $\sigma_{wf}$ will even improve, particularly for angled tracks. 
The $\sigma_{wf}$ can be even significantly lower than  $\sigma_{Lf}$ for LGAD detectors. 

In spite of somewhat shorter rise time, lack of sizable multiplication and 
consequently smaller $S/N$  means that jitter contribution to time resolution is 
relatively more important and possibly even dominant. For a typical rise time of $\sim 500$ ps a $S/N\sim20$ 
is required for $\sigma_j \sim \sigma_{wf}$, which for a 300 $\upmu$m thick silicon 3D detector translates 
to an equivalent noise charge of $ENC\sim 1000$ $e_0$. Keeping the noise small is 
therefore of utmost importance, probably excluding large pad detectors with small cell 
size due to its larger capacitance.  
The capacitance of a 3D detector is of the order $\sim 20$ pF/mm$^2$ for 50$\times$50 $\upmu$m$^2$ cells and a 
300 $\upmu$m thick detector, compared to around 2-3 pF/mm$^2$ for the 50 $\upmu$m thick LGAD detector.
Separation of the readout pad into smaller sub-pads with separate analog part and shared digital functionality may be 
one of the solutions.

For a given cell size configurations with more junction electrodes (2E) would have 
even smaller $\sigma_{wf}$, however, for small cell size devices the ratio of inactive 
(columns) to active part (bulk) of the detector would become larger. The same is true also 
for capacitance.

For an ideal detector being able to measure hit position as well as precise 
timing information a careful optimization of detection efficiency, noise occupancy 
and time resolution would be required and may yield a different design than
that for tracking detector only.

\section{Conclusions}

Timing performance of 3D detectors was simulated and measured with a 
single 50$\times$50 $\upmu$m$^2$ cell structure produced on a 300 $\upmu$m thick 
high resistivity wafer with a single 10 $\upmu$m wide $n^+$ readout column. 
The simulation results showed that for perpendicular tracks the timing resolution 
depends strongly on the cell size and that the dominant contribution to the time 
resolution is that of different induced current pulse shapes due to different hit positions. 
Unlike in LGADs, the Landau fluctuations, do not contribute much to the 
time resolution which allows the use of thick detectors.

The time resolution of a 3D detector with cell size 50$\times$50 $\upmu$m$^2$ cell (1E) 
is limited to around 45 ps for perpendicular tracks at 50 V and -20$^\circ$C for a single cell readout mode. 
For inclined tracks and multi-cell readout mode the minimum resolutions comparable 
or lower than that 26 ps due to Landau fluctuations in thin 50 $\upmu$m LGADs can be reached. 
The measurements performed at room temperature agreed well with simulations. Noise jitter of 47 
ps and time resolution of 75 ps were measured for $^{90}$Sr electrons in 300 $\upmu$m 
thick 3D detector at 50 V and room temperature. 

This is very promising as the 3D detectors, unlike LGADs, have 100\% fill factor and for small 
cell sizes exhibit very high radiation tolerance. Therefore the shift of operation  voltages to 
larger values after irradiation or even the onset of charge multiplication may even lead to 
significant improvement of their time resolution. The drawback is a higher capacitance which 
will increase the jitter and should be carefully optimized in terms of number of electrodes 
and thickness for the required performance. 

\section*{Acknowledgment}

Part of this work has been financed by the Spanish Ministry of Economy and Competitiveness through the Particle Physics National Program (FPA2015-69260-C3-3-R and FPA2014-55295-C3-2-R), by the European Union’s Horizon 2020 Research and Innovation funding program, under Grant Agreement no. 654168 (AIDA-2020). The work was also financed by the Slovenian Research Agency (ARRS) within the scope of research program P-00135.

\end{document}